\documentclass[]{llncs}
\usepackage{graphicx}
\pdfoutput=1
\usepackage[utf8x]{inputenc}
\usepackage{epstopdf}
\usepackage{algorithm,algorithmic}
\usepackage{amsmath,amssymb,amsfonts}
\usepackage{textcomp}
\usepackage{xcolor}
\begin{document}
\title{ A 2-Competitive Largest Job on Least Loaded Machine Online Algorithm based on Multi Lists Scheduling Model}
%
%
\author{Debasis Dwibedy  \and
Rakesh Mohanty}
\authorrunning{D. Dwibedy and  R. Mohanty}
%
\institute{Veer Surendra Sai University of Technology, Burla, 768018, India
\email{\{debasis.dwibedy,rakesh.iitmphd\}@gmail.com}}
%
\maketitle              
\begin{abstract}
Online scheduling in identical machines with makespan minimization has been a well studied research problem in the literature. In this problem, the scheduler receives a list of jobs one by one and assigns each incoming job on the fly  irrevocably to one of the machines before the arrival of the next job. In this paper, we study a variant of the online scheduling problem, where jobs are requested on the fly from $k$ distinct sources. On simultaneous arrival of jobs from multiple sources, it is a non-trivial research challenge to ensure fairness in scheduling with respect to distinct sources, while optimizing the overall makespan of the schedule. We develop a novel \textit{Multi Lists Scheduling Model(MLS)} and propose a $2$-competitive deterministic online algorithm namely \textit{Largest Job on Least Loaded Machine(LJLLM)} based on the \textit{MLS} model to address the above mentioned research challenge. 

\keywords{Competitive Analysis \and Identical Machines \and Multi Lists \and Makespan \and Online Scheduling \and Parallel Submissions.}
\end{abstract}
\section{Introduction}\label{sec:Introduction}
Design of algorithms with unavailability of complete information is a challenging research area in various domains of Computer Science. In most of the practical applications, we have to design algorithms in which inputs are available in an incremental fashion and decisions are made without knowledge of future inputs [1]. The algorithms designed for such applications are popularly known as \textit{online algorithms} [2, 3]. However, in traditional \textit{offline algorithms}, the whole set of inputs are available for processing by the algorithm at the outset.\\
\textbf{Practical Motivation.} In this paper, we consider the online non-preemptive scheduling of $n$ independent jobs on $m$-identical parallel machines, where $m \geq 2$ and $n>>>m$. We consider a realistic scenario, where jobs are submitted in parallel from $k$ distinct sources incrementally one by one such that each source can submit at most one job at a particular time step. The sources may be the users of multiprocessing systems, vendors ordering for products to a production company, links submitting packets to a central router or clients requesting for services to the cloud servers [4]. Here, the scheduler is constrained to  schedule a batch of $k$ incoming jobs, one from each source at each time step irrevocably before the arrival of the next batch with an objective to minimize the completion time of the job schedule i.e. makespan.\\
\textbf{Research Question.} On simultaneous arrival of jobs from multiple sources, the scheduler must ensure \textit{fairness} in scheduling decision making with respect to distinct sources. Generally, in an interactive multiprocessing environment, it is desirable to share the machines not only based on the jobs but also based on the distinct sources of job requests [5]. Therefore, we pose the following non-trivial  research question and attempt to address the solution.  \\
\textit{Can we design a model and efficient constant competitive online scheduling algorithm that explicitly schedules the simultaneous arrival of jobs from  distinct sources with an objective to minimize the makespan and ensures fairness in scheduling?}\\
According to our knowledge, there is no study in the literature regarding explicit ordering of the jobs with respect to distinct sources and scheduling of simultaneously arriving jobs. To address the above research question, we propose the \textit{Multi Lists Scheduling(MLS)} model, where we consider multiple parallel lists, one for each source of job requests. The objective is to provide simultaneous arrival of jobs to the scheduler at each time step so as to ensure fairness in scheduling with respect to distinct sources. We formally present the \textit{MLS} model as follows.
\subsection{Multi Lists Scheduling Model}\label{subsec:Multi List Scheduling Model} 
\begin{figure}[ht]
\centering
\includegraphics[scale=0.6]{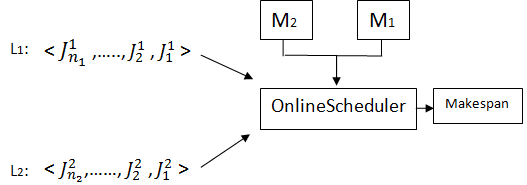}
\caption{MLS model with two Lists and two Machines}
\label{fig:Amodelofmultilistscheduling.png}  
\end{figure}

For simplicity and basic understanding of the readers, we present the \textit{MLS} model for online scheduling of $n$ jobs with two lists and two machines  as shown in \textit{Figure} \ref{fig:Amodelofmultilistscheduling.png}. Let $M_1$, $M_2$ be two identical machines and $L_1$, $L_2$ be two sub-lists representing two sources of incoming jobs with $n_1$, $n_2$ number of jobs respectively, where $L_1:<J^{1}_1, J^{1}_2,...,J^{1}_{n_1}>$
and $L_2:<J^{2}_1, J^{2}_2,...,J^{2}_{n_2}>$. The jobs from both lists $L_1$ and $L_2$ arrive one by one in parallel at time $t=0$. An online scheduler receives at most two jobs(one each from $L_1$ and $L_2$) at any time step $t$ and irrevocably schedules the jobs before the arrival of the next batch of jobs at time step $t_1$=$t+\epsilon$, where $\epsilon$ is a positive constant. The outcome of the scheduling is the earliest time by which all jobs complete their processing, otherwise known as \textit{makespan}.\\
We now define the \textit{MLS} model by extending the concept of $2$ machines and $2$ lists to $m$ machines and $k$ lists as follows. Let us consider a list $L$ of $n$ independent jobs is partitioned into  $k$ sub-lists(each sub-list is denoted by $L_r$)   representing $k$ distinct sources, where $L= \bigcup _{r=1}^{k}{L_r}$, and $L_x\cap L_y$=$\phi$, where $x\neq y$ and $1\leq (x, y)\leq k$. We have $|L_r|=n_r$, $1 \leq r \leq k$ and $\sum_{r=1}^{k}{n_r}=n$. We are given $m$-identical machines, where $2 \leq m << n$. The jobs arrive one by one simultaneously from all sub-lists $L_r$, where $1 \leq r \leq k$. So, at any time step $t_i$, exactly $r$ jobs are submitted by $r$ sub-lists to the scheduler and are irrevocably scheduled on the machines prior to the submissions of next $r$ jobs at time step $t_{i+1}$=$t_i+\epsilon$. The processing time of a job is known only upon its arrival. Scheduling decision is irrevocable with no preemption of jobs. The goal is to obtain a minimum makespan. 
\subsection{Research Motivation}\label{subsec:Research Motivation}
It has been proved that the $m$-machine scheduling problem is NP-Complete for $m \geq 2$ [6]. Finding the optimum schedule for $m=2$ is equivalent to the well-known \textit{partition} problem, where a weighted set is to be partitioned  in to two sets such that both sets have equal weights [15]. In case of our \textit{MLS} model, at most $n_1*n_2*....*n_{k-1}*n_k$ distinct job pairs can be available to the scheduler in ($n_1*n_2....*n_{k-1}*n_k$)! different ways. As a whole, we can schedule  $n$ jobs on $m$ machines in $m^{n}$ possible ways. In worst case, the probability of getting an optimum solution is $\frac{1}{{m}^{n}}$. Further, if the input jobs are available online from multiple sources, then it is challenging to design efficient online algorithms which could ensure fairness based on scheduling decision making with respect to distinct sources. The performance of such algorithms is evaluated generally by \textbf{Competitive analysis} method with \textit{Competitive Ratio(CR)} [7]. \textit{CR} is the  ratio between the objective values obtained by an online algorithm \textit{ONL} and the optimum offline algorithm \textit{OPT}. Let $C_{OPT}(I)$ and  $C_{ONL}(I)$ are the objective values incurred by \textit{OPT} and \textit{ONL } respectively in processing a sequence of inputs $I:<i_1, i_2,.....i_n>$.  Algorithm \textit{ONL} is $c$-competitive for the smallest $c$ and a positive constant $b$, iff ${C_{ONL}(I)} \leq c\cdot C_{OPT}(I)+b$, for all legal sequences of $I$, where $b, c \geq 1$. Here $c$ is referred to as \textit{competitive ratio}. A non-trivial challenge is to obtain a \textit{CR}  bounded by a constant and closer to $1$. 
\subsection{Our Contributions}\label{subsec:Our Contributions} 
We propose \textit{MLS} model to theoretically analyze the online scheduling of jobs submitted in parallel from $k$ distinct sources. The \textit{MLS} model inherently ensures fairness in scheduling by considering at each time step at most one job from each source of job request prior to make a scheduling decision.  
We propose a deterministic online algorithm, known as \textit{Largest Job on Least Loaded Machine(LJLLM)} based on the \textit{MLS} model for $m$-machines online scheduling of $n$ independent jobs submitted from $k$ distinct sources. At any time step, algorithm \textit{LJLLM} receives $r$ jobs, where $1 \leq r \leq k$ and assigns jobs one by one in non-increasing order of their processing times to the machine with minimum load after each assignment of a job. Here, every time a decision is made by considering $r$ distinct sources of job requests to ensure fairness in scheduling. The objective is to balance total load among $m$-machines so that the maximum load incurred on any of the machines would be minimized. We show that algorithm \textit{LJLLM} achieves an upper bound of $2$ on the competitive ratio for $m > 2 $.
\section{Background and Preliminaries}\label{sec:Background and Preliminaries}
In this section, we present the basic terminologies and notations related to our work as follows.
\subsection{Basic Terminologies and Notations}\label{subsec:Basic Definitions and Notations}
 Let $J^{r}_{i}$ be the $i^{th}$ job from the $r^{th}$ sub-list($L_r$), where $1 \leq r \leq k$ and $1 \leq i \leq n-k+1$.
We have at least one job and at most $n-k+1$ jobs in a sub-list. A sub-list can have $n-k+1$ jobs if and only if rest $k-1$ lists have exactly one job each. Let $p^{r}_{i}$ be the processing time of job $J^{r}_{i}$. We use the terms processing time and \textit{size} of a job in the same sense. The sub-lists of jobs with their processing times are presented as follows:  $L_1=<J^{1}_{1}/p^{1}_{1}, J^{1}_{2}/p^{1}_{2},.....J^{1}_{n_1}/p^{1}_{n_1} >$ and $L_k=<J^{k}_{1}/p^{k}_{1}, J^{k}_{2}/p^{k}_{2},.....J^{k}_{n_k}/p^{k}_{n_k} >$, where $1 \leq k \leq n$.  Let $M_j$ be the $j^{th}$ machine, where $1 \leq j \leq m$. The machines are \textit{identical} with respect to their speeds, i.e.  $p^{r}_{i}$ of any $J^{r}_{i}$ is same for all $M_j$. Let $l_j$ be the \textit{load} of any $M_j$, which is the total processing time of the jobs assigned to $M_j$ i.e. $l_j=\sum_{r}^{}\sum_{i}^{}{(p^{r}_{i})}$. The largest load incurred on any $M_j$ after the completion of the job schedule is known as makespan ($C_{max}$) and considered as the latest completion time among all $J^{r}_{i}$. We have $C_{max}= max\{l_j|1 \leq j \leq m\}$. \textit{Idle time} is the duration of time in which a machine $M_j$ does not process any $J^{r}_{i}$  and is presented as $\varphi$ in the timing diagram of a schedule.  $\varphi_{j}$ denotes the idle time of $M_j$ and the total idle time incurred by all $M_j$ can be: $\sum_{j=1}^{m}{(\varphi_{j})}$.
\subsection{Overview of Related Work}\label{subsec:Overview of Related Work}
Online scheduling problem has been well studied in the literature with multitude of variants [1, 8]. However, the multi user and interactive parallel systems generally support concurrent processing of multiple jobs which are requested from different users. Here, the primary objective is to ensure fair sharing of resources among the jobs based on the distinct users with maximization of throughput of the system [5]. In general, fairness in scheduling means quick response to each request and user satisfaction. Fair share scheduling has been studied in various offline settings (for a comprehensive review please see [9]).
\\
We now present a brief overview of related work on online scheduling and important known results as follows. Online $m$-machine scheduling problem has received significant research interest since the pioneering work of Graham in [10]. Graham introduced the famous \textit{List Scheduling(LS)} model for online scheduling. Here, jobs arrive online from a list in order and the objective is to obtain minimum time of completion for all jobs. He proposed the $2$-competitive \textit{List Scheduling algorithm(LSA)}. However, he did not consider specifically the sources of job requests and schedules an available job to the machine that has currently the minimum load. In [11], Graham considered the offline setting of \textit{LS} model, where a sequence of jobs are ordered in a list in non-increasing \textit{sizes}. For this setting, he proposed the algorithm \textit{Largest Processing Time(LPT)}, which follows the order of the list to schedule jobs on the free machines. He proved a worst case performance ratio of $1.33-\frac{1}{3m}$ for algorithm \textit{LPT}. There after, many improved online algorithms [12-20] have been proposed for the online $m$-machines scheduling problem to minimize the makespan. For this problem, we have the best upper bound of $1.9201$ on the competitive ratio due to Fleischer and Wahl [18]. The current best lower bound on the competitive ratio of $1.88$ is due to Rudin III [19]. \\
Recently, \textit{resource augmentation} policies have gained significant attention to improve the performance of online scheduling algorithms. Here, an online algorithm is given with some \textit{Extra Piece of Information(EPI)} on input jobs or  power in terms of  additional space while processing a sequence of job requests. One of the most considered \textit{EPI} in the literature [21-23] for obtaining improved competitive online scheduling algorithm is the "\textit{arrival order of the jobs}". The best competitive ratio of $1.25$ for scheduling of jobs arriving online in decreasing order of their sizes is due to Cheng et al. [22]. Many authors [24-28] have considered a buffer of certain length  $k$ for keeping $k$ incoming jobs before making a scheduling decision. Lan et al. [27] obtained the best competitive ratio of $1.5$ with a buffer of size $1.5m$ for any $m \geq 2$. Albers and Hellwig [29] proposed an ($1.33+\epsilon$)-competitive algorithm, where $0 < \epsilon \leq 1$. They considered extra space for the algorithm to maintain parallel schedules while processing a job sequence and finally choose the best schedule. \\
To the best of our knowledge, there is less attention and limited work in the current literature [30] for online scheduling in the context of parallel submissions with fairness as an objective.
\section{Our Proposed Online Algorithm LJLLM }\label{sec:Our Proposed Online Algorithm LJLLM }
\begin{figure}[!ht]
\centering
\includegraphics[scale=0.5]{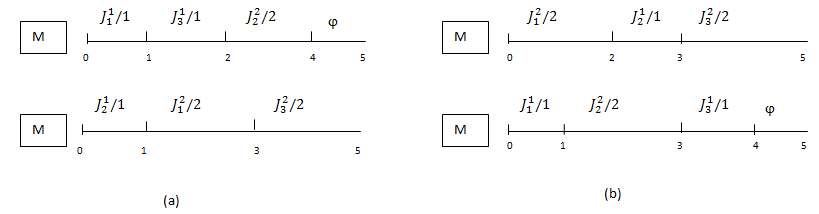}
\caption{(a) Online List Scheduling  (b)Online Multi Lists Scheduling}
\label{fig:onlinelistschedulingandfairsharelistscheduling.png}  
\end{figure}

We present the pseudocode of algorithm \textit{LJLLM} for $m$-machines($m \geq 2$), $n$ jobs and $k=2$ in \textit{Algorithm 1}.
\begin{algorithm}
\caption{LJLLM}
\begin{algorithmic}[1]
\STATE Initially, i=1, $l_1=l_2=.............l_m=0$\\
\STATE MIN()\\
\STATE \hspace*{0.2cm} BEGIN\\
\STATE \hspace*{0.5cm} $l_{min}=min\{l_j|j=1, 2,....m-1, m\}$\\
\STATE \hspace*{0.5cm} Select any $M_j$ for which $l_j=l_{min}$ \\
\STATE \hspace*{0.5cm} Index $M_j$ as $M_1$. \\
\STATE \hspace*{0.5cm} RETURN $M_1$\\
\STATE \hspace*{0.2cm} END\\
\STATE WHILE new jobs ($J^{1}_{i}$ AND $J^{2}_{i}$) arrives DO\\
\STATE \hspace*{0.2cm} BEGIN\\
\STATE \hspace*{0.5cm} CALL MIN()\\
\STATE \hspace*{0.5cm} IF ($p^{1}_{i} \geq p^{2}_{i}$) DO \\
\STATE \hspace*{0.8cm} Assign $J^{1}_{i}$ to $M_1$\\
\STATE \hspace*{0.8cm} UPDATE $l_1=l_1+p^{1}_{i}$\\
\STATE \hspace*{0.8cm} CALL MIN()\\
 \STATE \hspace*{0.8cm} Assign $J^{2}_{i}$ to $M_1$.\\
 \STATE \hspace*{0.8cm} UPDATE $l_1=l_1+p^{2}_{i}$\\
\STATE \hspace*{0.8cm} $i=i+1$ \\
\STATE \hspace*{0.5cm} END IF \\
\STATE \hspace*{0.5cm} ELSE DO \\
\STATE \hspace*{0.8cm} Assign $J^{2}_{i}$ to $M_1$\\
\STATE \hspace*{0.8cm} UPDATE $l_1=l_1+p^{2}_{i}$\\
\STATE \hspace*{0.8cm} CALL MIN()\\
\STATE \hspace*{0.8cm} Assign $J^{1}_{i}$ to $M_1$.\\
 \STATE \hspace*{0.8cm} UPDATE $l_1=l_1+p^{1}_{i}$\\
\STATE \hspace*{0.8cm} $i=i+1$ \\ 
\STATE \hspace*{0.5cm} END ELSE\\
\STATE \hspace*{0.2cm} END WHILE \\
\STATE WHILE new job ($J^{1}_{i}$ OR $J^{2}_{i}$) arrives DO\\
\STATE \hspace*{0.2cm} BEGIN\\
\STATE \hspace*{0.5cm} CALL MIN()\\
\STATE \hspace*{0.5cm} Assign $J^{1}_{i}$ to $M_1$ OR $J^{2}_{i}$ to $M_1$.\\
\STATE \hspace*{0.5cm} UPDATE $l_1=l_1+p^{1}_{i}$ OR $l_1=l_1+p^{2}_{i}$\\
\STATE \hspace*{0.5cm} $i=i+1$ \\  
\STATE \hspace*{0.2cm} END WHILE\\
\STATE Return \hspace*{0.3cm} $l_j=max\{l_j|j=1, 2,......m\}$
\end{algorithmic}
\end{algorithm}
Algorithm \textit{LJLLM} can further be extended for $k > 2$. Now, algorithm \textit{LJLLM} assigns the largest $J^{r}_{i}$, where $1 \leq r \leq k$ and $1 \leq i \leq n-k+1$ among the available jobs to the least loaded machine $M_1$. Then the load of each machine is updated so that the second largest $J^{r}_{i}$ can be assigned to $M_1$ of newly indexed machines and this continues iteratively till all $r$ jobs get scheduled in one slot. Similarly, algorithm \textit{LJLLM }takes at least $1$ slot and at most $n$ slots to schedule $n$ jobs. \\
\textbf{Significance of MLS model with Algorithm LJLLM.} To address the limitation of fair sharing of jobs with respect to distinct sources in the \textit{LS} model with algorithm \textit{LSA} as shown in \textit{Figure} \ref{fig:onlinelistschedulingandfairsharelistscheduling.png}(a), we present algorithm \textit{LJLLM} using \textit{MLS} model through an illustration as shown in \textit{Figure} \ref{fig:onlinelistschedulingandfairsharelistscheduling.png}(b). 
In \textit{Figure} \ref{fig:onlinelistschedulingandfairsharelistscheduling.png}, a special case is illustrated for both \textit{LS} and \textit{MLS} models. A large number of jobs from a single source are available upfront in the list in \textit{LS} model. However, this case can be avoided by \textit{MLS} model.
Let us consider $m=2$, $n=6$, $n_1=3$, $n_2=3$, arrival sequence of jobs for \textit{LS} model $L:<J^{1}_{1}/1,J^{1}_{2}/1,J^{1}_{3}/1,J^{2}_{1}/2,J^{2}_{2}/2,J^{2}_{3}/2>$, $L$ is partitioned into two lists for \textit{MLS} model as follows: $L_1:<J^{1}_{1}/1,J^{1}_{2}/1,J^{1}_{3}/1>$ and $L_2:<J^{2}_{1}/2,J^{2}_{2}/2,J^{2}_{3}/2>$. Let us consider, two jobs are available simultaneously, one from each list, in the order $\sigma: <(J^{1}_{1}/1, J^{2}_{1}/2), (J^{1}_{2}/1, J^{2}_{2}/2), (J^{1}_{3}/1, J^{2}_{3}/2)>$.
\section{Our Results on Competitive Analysis of Algorithm LJLLM}\label{sec:Our Results on Competitive Analysis of Algorithm LJLLM}
\textbf{Theorem 1.} \textit{Algorithm LJLLM is $2$-competitive, where $m > 2 $ \hspace* {0.1cm} and \hspace* {0.1cm} $1 \leq k < n$.} \\\\
\textit{Proof:}  Let us consider a list $L= \{J^{1}_{1}, J^{1}_{2}........J^{k}_{n}
\}$ of $n$ jobs that are partitioned into $k$ parallel lists such that the number of jobs $n_{r}$ in any $k$ is $1 \leq n_{r} \leq (n-k+1)$ and $\sum_{r=1}^{k}{(n_r)}=n$. The total load incurred by $n$ jobs is $\sum_{r=1}^{k} \sum_{i=1}^{}({p^{r}_{i}})$. W.l.o.g, for simplicity in analysis we consider $J^{r}_{i}$ as $J_{i}$ and $p^{r}_{i}$ as $p_{i}$, where $1 \leq i \leq n$. Therefore, We have \\\\
\hspace*{4.6cm} $\sum_{r=1}^{k} \sum_{i=1}^{}{p^{r}_{i}}$=$\sum_{i=1}^{n}{p_{i}}$ \hspace*{3.2cm} (1)\\\\
Let $J_{b}$ be the largest job with \textit{size} $p_{b}$, where, $p_{b}=\max_{1 \leq i \leq n}{p_i}$. Let the optimal makespan obtained by \textit{OPT} for $L$ be $C_{OPT}(L)$ and makespan incurred by algorithm \textit{LJLLM} be $C_{LJLLM}(L)$.\\\\
\textit{We now provide the computation of OPT:} Let us consider the following general bounds [1] for computation of \textit{OPT} \\\\
\hspace*{4.6cm} $C_{OPT}(L)\geq \frac{1}{m} \sum_{i=1}^{n} {(p_{i})}$. \hspace*{3.0cm} (2)\\\\
The following inequality is obvious. \\
\hspace* {4.6cm} $C_{OPT}(L) \geq p_{b}$. \hspace* {4.6cm} (3)\\\\
Further, we prove this theorem by considering different cases which are presented in lemma 1.1 and 1.2  as follows.\\\\
\textbf{Lemma 1.1.} \textit{LJLLM is ($2-\frac{1}{m}$)-competitive, where each list $L_r$ has $n_r$ number of jobs, where $1 \leq n_r \leq (n-k+1)$.}\\\\
\textit{Proof:} A machine $M_j$ becomes idle, when it completes the execution of all jobs assigned to it and there is no unscheduled job left for its processing. Therefore the following inequality holds.\\\\
\hspace* {4.6cm} $\sum_{j=1}^{m}{(\varphi_j)} \leq (m-1)p_b$ \hspace*{3.6cm} (4)\\ \\
Makespan obtained by algorithm \textit{LJLLM} for $k$ sequences of total $n$ jobs is equivalent to the largest load incurred on any of the machines. Therefore, we have \hspace* {0.2cm} $C_{LJLLM}(L) \leq \max_{1 \leq j \leq m}{(l_j)}$ this implies\\ 
\hspace* {0.7cm} $C_{LJLLM}(L) \leq \frac{1}{m}(\sum_{i=1}^{n}{p_i}+\sum_{j=1}^{m}{\varphi_{j}})$\\
By (2) and (4), we have\\
\hspace* {0.7cm} $C_{LJLLM}(L) \leq C_{OPT}(L)+ \frac{1}{m}(m-1)p_b$\\
By (3), we have \\
\hspace* {0.7cm} $C_{LJLLM}(L) \leq C_{OPT}(L)+ \frac{1}{m}(m-1)C_{OPT}(L)$\\
\hspace* {0.7cm} $C_{LJLLM}(L) \leq (2-\frac{1}{m})C_{OPT}(L)$ \hfill\(\Box\) \\\\
\textbf{Lemma 1.2.} \textit{LJLLM is ($2-\frac{1}{m}$)-competitive for $k$ parallel lists each with equal number of jobs}.\\\\
\textit{Proof:}  Let's consider that each list $L_r$, $1 \leq r \leq k$, has ($\frac{n}{k}$) jobs respectively. W.l.o.g,  we assume that at the time of arrival of last $k$ jobs, ($n-k$) jobs have already been scheduled. At this time, let $l_1=\min\{l_j|1 \leq j \leq m\}$, then we have the following inequality \\
\hspace* {4.6cm}  $l_1 \leq \frac{1}{m} \sum_{i=1}^{n-k}{(p_i)}$ \hspace*{4.2cm} (5)\\\\
\textbf{Case 1 \{$k \leq m$\}:}\\ 
Let $J_x$, where $(n-k+1) \leq x \leq n$ has processing time of $p_x$ such that $p_x=\max\{p_i|(n-k+1) \leq i \leq n\}$ and if $p_x=p_b$. Then, by (5) we have \\
\hspace* {4.4cm} $C_{LJLLM}(L) \leq [\frac{1}{m} \sum_{i=1}^{n-k}{(p_i)}]+p_x$ \hspace* {2.0cm} (6)\\\\
By the equal load sharing nature of \textit{OPT} and if $p_x=p_b$, further (6) can be:\\
\hspace* {0.7cm} $C_{LJLLM}(L) \leq C_{OPT}(L)-\frac{p_b}{m}+p_b \leq C_{OPT}(L)- \frac{C_{OPT}(L)}{m}+C_{OPT}(L)$\\
 \hspace* {0.7cm} $C_{LJLLM}(L) \leq (2-\frac{1}{m})C_{OPT}(L)$\\
\textbf{Case 2 \{$k > m$\}:}\\
 \textbf{ Case 2.1. \{$p_x \geq \frac{1}{m-1} \sum_{i=n-k+2}^{n}({p_i})$\}}:\\
 By (5) we have \\
\hspace* {0.7cm} $C_{LJLLM}(L) \leq [\frac{1}{m} \sum_{i=1}^{n-k}{p_i}]+p_x$\\
 Following the derivation of case 1. we have \\
\hspace* {0.7cm} $C_{LJLLM}(L) \leq (2-\frac{1}{m}).C_{OPT}(L)$.\\\\
 \textbf{ Case 2.2  \{$p_x < \frac{1}{m-1} \sum_{i=n-k+2}^{n}{(p_i)}$\}}:\\
Suppose, job $J_t$ with processing time of $p_t$ completes at time $T$ and let we have $C_{LJLL}(L) \leq T$.  Let $T-p_t=t$(Arrival time of $J_t$). Therefore, the following inequality holds.\\
\hspace* {0.7cm} $t \leq \frac{1}{m}\sum_{i=1}^{n-k}{(p_i)}+\frac{1}{m}\sum_{i=n-k+1}^{n-1}{(p_i)}$, further, we have\\
 \hspace* {0.7cm} $t+p_t \leq \frac{1}{m}\sum_{i=1}^{n-1}{(p_i)}+p_t$\\
 Now, if $p_t=p_b$, then we have\\
 \hspace* {0.7cm} $t+p_b \leq \frac{1}{m}\sum_{i+1}^{n-1}{(p_i)}+p_b$\\
 \hspace* {0.7cm} $C_{LJLLM}(L) \leq \frac{1}{m}\sum_{i=1}^{n-1}{(p_i)}+p_b \leq \frac{1}{m}[\sum_{i=1}^{n-1}{(p_i)}+p_b]+(\frac{m-1}{m})p_b$\\
 By (2) and (3), we have\\
 \hspace* {0.7cm} $C_{LJLLM}(L) \leq C_{OPT}(L)+(\frac{m-1}{m})C_{OPT}(L)$\\
 \hspace* {0.7cm} $C_{LJLLM}(L) \leq (2-\frac{1}{m})C_{OPT}(L)$\\\\
Lemmas 1.1 and  1.2 can obtain the following inequality for $m \rightarrow \infty$ \\
 \hspace* {0.7cm} $C_{LJLLM}(L) \leq 2.C_{OPT}(L)$ \hfill\(\Box\)\\\\
\textbf{Corollary 1.3.} \textit{Algorithm LJLLM is $2$-competitive, where all jobs are of unit size.}\\\\
\textbf{Lemma 1.4.} \textit{LJLLM is $1.33$-competitive, where $k=n$.}\\\\
\textit{Proof:} Let's denote $p^{r}_{i}$, where $1 \leq r \leq k$ and $1 \leq i \leq n-k+1$ as $p_i$, where $1 \leq i \leq n$. Let's update the indices of the jobs as $1,2,...n$ and organize the jobs in a list $L$ such that $p_1 \geq p_2 \geq ....\geq p_n$. Then, \textit{LJLLM} schedules the jobs one by one in the following order $\sigma: <J_1, J_2,....J_{n-1}, J_n>$ on the machine which has minimum load after the assignment of each job. Let the costs obtained by \textit{LJLLM} and \textit{OPT} be $C_{LJLLM}(L)$ and $C_{OPT}(L)$ respectively. From Theorem. 2 of Graham [11], we have:\\
\hspace* {4.6cm} $C_{LJLLM}(L) \leq (\frac{4}{3}-\frac{1}{3m})C_{OPT}(L)$. \hspace* {1.8cm} (7)\\
For $m \rightarrow \infty$, (7) can be\\
\hspace* {4.6cm} $C_{LJLLM}(L) \leq (1.33)C_{OPT}(L)$. \hfill\(\Box\)\\\\
\textbf{Lemma 1.5.} \textit{LJLLM is $1$-competitive with MLS model, where $m=1$ or $n \leq m$.}\\\\
\textit{Proof:} It is quite obvious that for $m$=$1$, both \textit{OPT} and \textit{LJLLM} obtain makespan \\  
\hspace* {4.6cm} $C_{OPT}(L)$=$C_{LJLLM}(L) \geq \sum_{r=1}^{k}\sum_{i}^{}{(p^{r}_{i})}$ \hspace* {0.9cm} (8)\\\\
Let $p_b$=$\max \{p^{r}_{i}|1 \leq r \leq k\hspace* {0.3cm}and \hspace* {0.3cm} i \geq 1\}$. If $n \leq m$, then \textit{OPT} and \textit{LJLLM} obtain makespan\\ 
\hspace* {4.6cm} $C_{OPT}(L)$=$C_{LJLLM}(L) \geq p_b$ \hspace* {2.7cm} (9)\\
From equations (8) and (9), it is clear that algorithm \textit{LJLLM} is $1$-competitive. \hfill\(\Box\)
\section{Conclusion and Future Scope}\label{sec:Concluding Remarks}
To address the challenging issue of handling incoming jobs from multiple sources for online scheduling on identical machines, we introduced the \textit{MLS} model. To achieve fairness in scheduling, we proposed the deterministic online scheduling algorithm \textit{LJLLM}. We analyzed the performance of algorithm \textit{LJLLM} by competitive analysis method and obtained an upper bound of $2$ on the competitive ratio. \\
\textbf{Future Scope.} It is an open problem to define fairness as a formal quantitative measure  for the analysis of performance of online scheduling algorithms. We believe that our study on \textit{MLS} model would possibly open up a new research direction in online scheduling for study of fairness as a performance measure. Practitioners can also characterize the inputs of any real world applications on the \textit{MLS} model and can analyze the performance of algorithm \textit{LJLLM}. In \textit{MLS} model, it will be interesting to compute the flow time for each job and to minimize the overall flow time for each sources of job requests and design efficient new fair online scheduling algorithms.
%
%
%

\begin{thebibliography}{}


\bibitem{Albers2009}
Albers S.(2009) "Online scheduling: In Introduction to Scheduling," \textit{edited by Y. Robert and F. Vivien. Chapman and Hall/CRC Press}, pp. 57-84.
\bibitem{Borodin1998}
Borodin A., and El-Yaniv R.(1998) "Online Computation and Competitive Analysis," \textit{Cambridge University press, Cambridge}.
\bibitem{Albers2003}
Albers S.(2003) "Online Algorithms: A Survey," \textit{Mathematical Programming}, \textbf{97}(1):3-26.
\bibitem{Albers1997}
Albers S.(1997) "Competitive online algorithms," \textit{OPTIMA: Mathematical Programming Society Newsletter}, \textbf{54}(1):1-8.
\bibitem{Kay1988}
Kay J., and Lauder P.(1988) "A fair share scheduler," \textit{Communications of the ACM}, \textbf{31}(1):44-55.
\bibitem{Garey1979}
Garey M. R., and Johnson D. S.(1979) "Computers and Intractability: A Guide to the Theory of NP-Completeness," $1^{st}$ Edition, \textit{Freeman}, 1979.
\bibitem{Sleator1985}
Sleator D. D., and Tarjan R. E.(1985), "Amortized efficiency of list update and paging rules," \textit{Communications of the ACM},  \textbf{28}(1):202-208.
\bibitem{Brucker2006}
Brucker P.(2006) "Scheduling Algorithms," $5^{th}$ Edition, \textit{springer}.
\bibitem{Feitelson1997}
Feitelson D. G.(1997) "Job Scheduling in Multi-programmed
Parallel Systems(Extended Version)," \textit{IBM Research Report}, \textbf{RC} 19790 (87657), Second Revision.
\bibitem{Graham1966}
Graham R. L.(1966) "Bounds for certain multiprocessor anomalies," \textit{Bell System Technical Journal}, \textbf{45}(1):1563-1581.
\bibitem{Graham1969}
Graham R. L.(1969), "Bounds on multiprocessor timing anomalies," \textit{SIAM Journal on Applied Mathematics}, \textbf{17}(2):416-429.
\bibitem{Bartal1992}
Bartal Y., Fiat A., Karloff H., and Vohra R.(1992) "New algorithms for an ancient scheduling problem," \textit{In Proceedings of the $24^{th}$ ACM Symposium on the Theory of Computing}, Victoria, Canada, pp. 51-58.
\bibitem{Galambos1993}
Galambos G., and Woeginger G. J.(1993) "An online scheduling heuristic with better worst case ratio than Graham's list scheduling," \textit{SIAM Journal on Computing}, \textbf{22}(2):349-355.
\bibitem{Karger1996}
Karger D. R., Phillips S. J., and Torng E.(1996) "A better algorithm for an ancient scheduling problem," \textit{Journal of Algorithms},\textbf{20} article no:19, pp. 400-430.
\bibitem{Bartal1994}
Bartal Y., Karloff H., and  Rabani Y.(1994) "A better lower bound for online scheduling," \textit{Information Processing Letters}, \textbf{50}(1):113-116.
\bibitem{Chen1994}
Chen B., Vliet A. V., and Woeginger G. J.(1994) "New lower and upper bound for online scheduling," \textit{Operation Research Letters}, \textbf{16}(1):221-230.
\bibitem{Albers1999}
Albers S.(1999) "Better bounds for Online scheduling," \textit{SIAM Journal on Computing}, \textbf{29}(1):459-473. 
\bibitem{Fleischer2000}
Fleischer R., and Wahl M.(2000) "Online scheduling revisited," \textit{Journal of Scheduling}, \textbf{3}(1):343-353.
\bibitem{Rudin III2001}
Rudin III J. F.(2001) "Improved bounds for the on-line scheduling problem," \textit{Ph.D. Thesis. The University of Texas at Dallas}, May.
\bibitem{Englert2008}
Englert M., Ozmen D., and Westermann M.(2008), "The power of reordering for online minimum makespan scheduling," \textit{In Proceedings of the $49^{th}$ Annual IEEE Symposium on Foundations of Computer Science}. 
\bibitem{Seiden2000}
Seiden S., Sgall J., and Woeginger G.(2000) "Semi-online scheduling with decreasing job sizes," \textit{Operations Research Letters},\textbf{27}(1):215-221.
\bibitem{Cheng2012}
Cheng T. C. E., Kellerer H., and Kotov V.(2012), "Algorithms better than LPT for semi-online scheduling with decreasing processing times," \textit{Operations Research Letters},\textbf{40}(1):349-352.
\bibitem{Tang2015}
Tang F., and Nie J.(2015) "LS algorithm for semi-online scheduling jobs with non-decreasing processing times," \textit{International Conference on Computers, Information system and Industrial Applications(CISIA)}.
\bibitem{Kellerer1997}
Kellerer H., Kotov V., Speranza M. G., and Tuza Zs.(1997) "Semi-online algorithms for the partition problem," \textit{Operations Research Letters}, \textbf{21}(1):235-242.
\bibitem{Zhang1997}
Zhang G.(1997) "A simple semi-online algorithm for $P_2 // C_{max}$ with a buffer," \textit{Information Processing Letters}, \textbf{61}(1):145-148.
\bibitem{Dosa2004}
Dosa G., and He Y.(2004) "Semi-online algorithms for parallel machine scheduling problems," \textit{Computing}, \textbf{72}(1):355-363.
\bibitem{Lan2012}
Lan Y., Chen X., Ding N., Dosa G., and Han X.(2012) "Online makespan scheduling with a buffer," \textit{Frontiers in Algorithms and Aspects in Information and Management}, pp. 161-171.
\bibitem{Chen2013}
Chen X., Xu Z., Dosa G., Han X., and Jiang H.(2013) "Semi-online hierarchical scheduling problems with buffer or re-arrangements," \textit{Information Processing Letters}, \textbf{113}(1):127-131.
\bibitem{Albers2016}
Albers S., and Hellwig M.(2016) "Online makespan minimization with parallel schedules," \textit{Algorithmica}, DOI 10.1007/s00453-016-0172-5.
\bibitem{Pinheiro2014}
Pinheiro V. G.(2014), "The management of multiple submissions in parallel systems:the fair scheduling approach," \textit{Ph.D. Thesis, Institute of Mathematics and Statistics, University of Sao Paulo, Brazil}, April.
\end{thebibliography}
%

\end{document}